\documentclass[twocolumn,superscriptaddress,preprintnumbers,pra]{revtex4}
\usepackage{epsfig}
\usepackage{amssymb}
\usepackage{amsmath}
\usepackage{epstopdf}
\usepackage{graphicx,color}  
\usepackage{bm}  
\usepackage{epstopdf}

\begin{document}

\title{Mean-square radii of two-component three-body systems in two
  spatial dimensions} 

\author{J.~H. Sandoval} 
\affiliation{Instituto
  de F\'\i sica Te\'orica, UNESP - Univ Estadual Paulista,
  C.P. 70532-2, CEP 01156-970, S\~ao Paulo, SP, Brazil}

\author{F.~F. Bellotti} 
\affiliation{Department of Physics and
  Astronomy, Aarhus University, DK-8000 Aarhus C, Denmark}

\author{A.~S. Jensen} 
\affiliation{Department of Physics and
  Astronomy, Aarhus University, DK-8000 Aarhus C, Denmark}

\author{M.~T. Yamashita} 
\affiliation{Instituto de F\'\i sica
  Te\'orica, UNESP - Univ Estadual Paulista, C.P. 70532-2, CEP
  01156-970, S\~ao Paulo, SP, Brazil} 

\date{\today }

\begin{abstract}
We calculate root-mean-square radii for a three-body system confined
to two spatial dimensions and consisting of two identical bosons ($A$)
and one distinguishable particle ($B$).  We use zero-range two-body
interactions between each of the pairs, and focus thereby directly on
universal properties.  We solve the Faddeev equations in momentum
space and express the mean-square radii in terms of first-order
derivatives of the Fourier transforms of densities.  The strengths of
the interactions are adjusted for each set of masses to produce equal
two-body bound-state energies between different pairs.  The mass
ratio, ${\cal A}=m_B/m_A$, between particles $B$ and $A$ are varied from
$0.01$ to $100$ providing a number of bound states decreasing from $8$
to $2$.  Energies and mean-square radii of these states are analyzed
for small ${\cal A}$ by use of the Born-Oppenheimer potential between
the two heavy $A$-particles.  For large ${\cal A}$ the radii of the
two bound states are consistent with a slightly asymmetric three-body
structure.  When ${\cal A}$ approaches thresholds for binding of the
three-body excited states, the corresponding mean-square radii diverge
inversely proportional to the deviation of the three-body energy from
the two-body thresholds.  The structures at these three-body
thresholds correspond to bound $AB$-dimers and one loosely bound
$A$-particle.
\end{abstract}

\maketitle

\section{Introduction}

The last decade has provided unprecedented accurate techniques to tune
the effective interactions between pairs of ultracold atoms in
extremely controllable external fields.  The effective two-body
interactions can be varied, for a number of special atoms, from
strongly attractive to strongly repulsive by use of the Feshbach
resonance technique, see review in Ref. \cite{pethick}.  The two-body
interactions are experimentally determined functions of the applied
magnetic field strength.  The atomic gases, possibly with different
atoms, are confined by flexible deformed external fields.  Ultimately
this allows extreme asymmetry corresponding to spatial dimensions
lower than three \cite{modugnoPRA03,gunterPRL05}.  Structures of
different systems can then be simulated and studied.

Our interests in the present investigation are three-body structures
and two-dimensional universal properties.  First, few-body physics is
by definition accurately solvable without approximations in contrast
to many-body physics. But now experimental tests can be made for many
of the claims derived by theoretical calculations.  We focus here on
the absolutely simplest unsolved system of three particles in two
spatial dimensions (2D).  Second, the physics in 2D differ enormously
from the much more known properties in three dimensions (3D) and for
that matter any other spatial dimension
\cite{castinpra2012,bellottifbs2015}.  Third, the many universal
properties are arguably the most important results confirmed by cold
atomic gas experiments.  Properly formulated they are applicable
throughout physics as independent of scale and details of the
corresponding potentials \cite{jensenrmp2004,braatenpr2006}. 

The third of these points refer to the concept of universality which
means that the related properties can be described by only a few
scale-parameters \cite{fre12}.  The interactions
between neutral atoms in dilute atomic gases are of very short range
compared to the two-body scattering length
\cite{kunitskiscience2015,cencekjcp2012,kolganovafbs2011}.  This
implies that low-energy observables are universal and only depend on
the scattering length which can be the same for disparate potentials.
Thus the universal regime can be defined as properties depending only
on lengths far larger than the range of the potential and far smaller
than the scattering length.  Thus, a zero-range interaction with its
strength proportional to the scattering length is automatically
focussing on universal properties.

The special interest in two dimensions is due to the properties
arising from the negative centrifugal barrier corresponding to the
relative coordinate between two particles.  This means that binding is
achieved with even an infinitesimal attraction between the two
particles.  Also three particles are bound in 2D with almost vanishing
attractive two-body potentials.  For identical bosons in the universal
regime all observables in 2D can be expressed as functions of only one
two-body scale parameter, e.g. the two-body scattering length
\cite{jensenrmp2004,braatenpr2006}.  In 3D, both a two- and a
three-body scale parameter are needed to describe universal observables
\cite{skorniakov1957}.  This difference is closely related to the
appearance of the pathological Efimov effect for three particles in 3D
\cite{efimov} and its absence in 2D \cite{nielsenpr2001}.  This
apparent discontinuity can now be studied with external traps varying
continuously from spherical to cylindrical geometry \cite{yama15}.

The status for weakly bound three-body systems in 2D is that energies
are well studied theoretically
\cite{bruchpra1979,adhikarijp1986,adhikaripra1993,nie99,bellottijpb2011,
  bellottipra2012,bellottijpb2013,khurifbs2002,levinsenprx2014,esry2014,esry2015}.
The simplest reference system we have is three identical bosons with
only two three-body bound states with energies, $E_3$, proportional to
the two-body energy, $E_2$, that is $E_3=16.52E_2$ and $E_3=1.27E_2$
\cite{bruchpra1979}.  The previous studies include energies of three
non-identical particles where different masses and scattering lengths
substantially complicate systematic characterization of the universal
properties \cite{nie99}.  The simplest asymmetric system, denoted
$AAB$, is formed by two identical bosons, $A$, and a distinguishable
particle, $B$.  With spin-independent interactions, the results also
apply to two identical fermions with symmetric spatial wave function.
The number of three-body bound states increases as the mass ratio,
${\cal A}=m_B/m_A$, between $B$ and $A$ decreases \cite{bellottipra2012}.
Then the $B$-particle can be exchanged more easily between the heavy
$A$-particles, which in turn generates an effective potential
eventually of infinite attraction for vanishing ${\cal A}$
\cite{limZPA80,bellottijpb2013,levinsenprx2014}.

In contrast to the energies, knowledge about structures of asymmetric
three-body systems in 2D is virtually not existing.  One reason is
that experiments have only now become feasible, but the lack of the
necessary operating theoretical techniques is probably also partly
responsible for this delay.  Most of the studies have employed
zero-range interactions treated in momentum space, and have calculated
energies without direct access to the corresponding wave
functions. The structures therefore require a substantially larger
additional effort.  Coordinate space calculations in 2D with direct
link to wave functions are also not available at the moment.

We shall continue to use the zero-range interaction with the easy
interpretation in terms of universal properties
\cite{yamashitapra2003}.  It is worth mentioning that it would be 
interesting to extend the present calculation to more than three 
atoms to see how the universal relations derived in \cite{hammerprl2004} 
would change for a mass-imbalanced situation.  We use tedious mathematics 
to express various radial moments of relative distances as derivatives of
momentum space integrals.  This is instead of the extensive numerical
computation of momentum-space wave functions with subsequent Fourier
transformation and derivation of structural properties.  We shall be
content with second radial moments, which are the simplest observable
quantities that carry structure information.  If necessary, higher
moments can also be computed by use of the same technique, but we
believe the second radial moments of various relative distances are
sufficient to extract the dominating underlying structure.

The paper is organized as follows. In section \ref{formalism} we give
first the appropriate Faddeev equations in momentum space for an $AAB$
system of one distinguishable and two identical particles.  Then we
construct the three-body wave functions and the form factors from
which the radii of interest are calculated.  In section \ref{mass} we
present and analyze numerical results of energy spectra and
mean-square radii, as functions of the mass ratio.  In particular, we
discuss the threshold behavior when a level is passing into the
three-body continuum of one particle and a bound pair.  Finally,
conclusions and perspectives are summarized in section
\ref{conclusions}.

\section{Theoretical formulation}
\label{formalism}

We consider a three-body system, $AAB$, consisting of two identical
bosons, $A$, of mass, $m_A$, and bound state energy, $E_{AA} < 0$, and
a third distinguishable particle, $B$, of mass, $m_B$, bound to each
$A$-particle with energy, $E_{AB} < 0$.  We denote the identical
particles by $A$ and $A'$ when it is necessary to distinguish their
coordinates.  The two-body interactions are assumed to be of very
short range, and in this paper they are parametrized and applied in the extreme
zero-range limit where only $s$-waves are important.  This rather
schematic interaction extracts universal properties due to the absence
of spatial regions inside the potential.

\subsection{Faddeev equations}

The preferred method in most investigations is the use of the Faddeev
decomposition in momentum space.  The theoretical formulation is a
little elaborate but available in details in the literature
\cite{braatenpr2006,fre12}.  We shall here only
specify our notation and present the Faddeev equations in momentum
space.  We shall then proceed to sketch the novel derivation of the
different mean-square radii expressed as derivatives of integrals over
the momentum-space solutions.  The three-body state, $ | \Psi_{AAB}
\rangle$, corresponds to the three-body momentum-space wave function,
$\Psi_{A'B,A}$, when it is expressed in terms of $\vec{p}_A$ and
$\vec{q}_A$, that are, respectively, the relative momenta between the $B$
and the $A'$-particle, and the center-of-mass of $A'+B$ and the
$A$-particle.  The Faddeev decomposition then amounts to
\begin{eqnarray}
\label{psiab}
&&\Psi_{AB,A}(\vec{q}_A,\vec{p}_A)\equiv \langle\vec{q}_A,\vec{p}_A\,|\Psi_{AAB}\rangle \\
\nonumber
&&=\frac{\chi_A(|\vec{q}_A|)+\chi_B
(|\vec{p}_A-\frac{{\cal A}}{{\cal A}+1}\vec{q}_A|)+\chi_A(|\vec{p}_A+
\frac{1}{{\cal A}+1}\vec{q}_A|)}{|E_3|+\frac{{\cal A}+1}{2{\cal A}}
\frac{p^2_A}{m_A}+\frac{{\cal A}+2}{2({\cal A}+1)}\frac{q^2_A}{m_A}},
\end{eqnarray}
where $|E_3|$ is the three-body binding energy, the mass ratio between
particles $B$ and $A$ is denoted ${\cal A}\equiv m_B/m_A$, and the
so-called spectator functions, the Faddeev components $\chi_A$ and
$\chi_B$, are given as functions of the relative momenta in the
corresponding Jacobi coordinates, that are, $\vec{q}_A$, $\vec{q}_B$ and
$\vec{q}_{A'}$, where
\begin{eqnarray}
\vec{q}_B = \vec{p}_A-\frac{{\cal A}}{{\cal A}+1}\vec{q}_A \;\;,\;\;
\vec{q}_{A'} = \vec{p}_A+ \frac{1}{{\cal A}+1}\vec{q}_A \; .
\end{eqnarray}

The spectator functions, $\chi_A$ and $\chi_B$, obey coupled integral
equations obtained after $s$-wave projection.  The detailed standard
derivation of these equations in 2D can be found in
\cite{bellottijpb2011}, with the result
\begin{eqnarray}
\label{chib}
\nonumber
&&\chi_B(q)=2\tau_{AA}\left(|E_3|+\frac{{\cal A}+2}{4{\cal A}}\frac{q^2}{m_A}\right)\\
&&\times \int d^2p G_1(p,q;E_3)\chi_A(p)\\
\label{chia}
&&\chi_A(q)=\tau_{AB}\left(|E_3|+\frac{{\cal A}+2}{2({\cal A}+1)}\frac{q^2}{m_A}\right)\\
\nonumber
&&\times\int d^2p
\Bigg[G_1(q,p;E_3)\chi_B(p)+G_2(q,p;E_3)\chi_A(p)\Bigg]. \\
&&G_1(q,p;E_3)=\frac{1}{|E_3|+\frac{q^2}{m_A}+\frac{{\cal A}+1}{2{\cal A}}\frac{p^2}{m_A}
+\frac{\vec{q}\cdot\vec{p}}{m_A}}  ,\\
&&G_2(q,p;E_3)=\frac{1}{|E_3|+\frac{{\cal A}+1}{2{\cal A}}\frac{(q^2+p^2)}{m_A}+\frac{1}{{\cal A}}
\frac{\vec{q}\cdot\vec{p}}{m_A}} ,
\end{eqnarray}
where the two-body T-matrices, $\tau_{A\alpha}$, of the interacting pair are
given by
\begin{equation}
\label{2btmatrix}
\tau_{A\alpha}(E)=\left[-4\pi\frac{m_Am_\alpha}{m_A+m_\alpha}
\ln\left(\sqrt{\frac{|E|}{|E_{A\alpha}|}}\right)\right]^{-1},
\end{equation}
where $E_{A\alpha}$ is the two-body binding energy of the pair
$A\alpha$, with $\alpha=A,B$.

\begin{figure}[htb!]
\epsfig{file=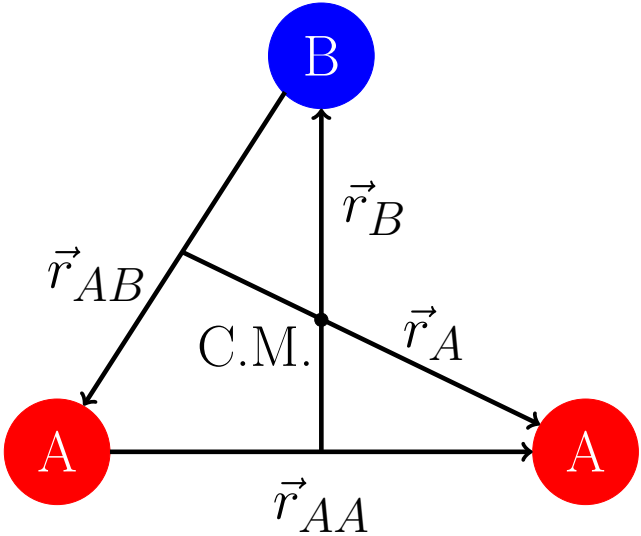,width=\columnwidth}
\caption{Schematic figure showing the three-body system of two
  identical bosons, $A$, and one distinguishable particle, $B$.  The
  vectors $\vec{r}_{AB}$ and $\vec{r}_{AA}$ are relative distances
  between the respective pairs of particles, whereas $\vec{r}_{A}$ and
  $\vec{r}_{B}$ denote distances between the specified particle and
  the center-of-mass of the remaining pairs.  The point marked
  C.M. means center-of-mass of the three-body system.  The notation
  used in the paper for the relevant distances are also shown.  }
\label{figradii}
\end{figure}

\subsection{Mean-square radii}

The distances in the three-body system can be seen in
Fig.\ref{figradii}, where $\vec{r}_{AB}$, $\vec{r}_{AA'}$ and
$\vec{r}_{A'B}$, are relative distances between pairs of particles,
and $\vec{r}_{A'}$, $\vec{r}_{A'}$ and $\vec{r}_B$ denote distances
between the center-of-mass of such pairs and the remaining last
particle.  The momenta, $\vec{q}_A$ and $\vec{p}_A$, are repectively
the canonically conjugated to the radii, $\vec{r}_A$ and
$\vec{r}_{AB}$.

The mean-square radii can be computed from the momentum-space wave
function by Fourier transform and subsequent calculation of the
corresponding matrix element.  To be precise we first define one-body
densities related to one set of Jacobi coordinates, $\vec{r}_A$ and
$\vec{r}_{AB}$, that is
\begin{eqnarray}
 \rho(\vec{r}_A)=\int d^2r_{AB} |\langle\vec{r}_A,\vec{r}_{AB}|
 \Psi_{AAB}\rangle|^2  \;,  \label{rhoa} \\  \label{rhoab}
 {\bar \rho}(\vec{r}_{AB})=\int d^2r_A |\langle\vec{r}_A,\vec{r}_{AB}|
 \Psi_{AAB}\rangle|^2    \; ,
\end{eqnarray}
where $\langle\vec{r}_A,\vec{r}_{AB}| \Psi_{AAB}\rangle$ is the
coordinate-space wave function expressed in terms of $\vec{r}_A$ and
$\vec{r}_{AB}$.  We first expand the Fourier transform, $F_A$ of
$\rho(\vec{r}_A)$ to second order in the corresponding momentum, that
is
\begin{eqnarray}
F_A(Q^2) &=& \int d^2r_A e^{\frac{i}{\hbar}\vec{Q}\cdot\vec{r}_A}\rho(r_A)
\approx  1-\frac{1}{4\hbar^2}Q^2 \langle r_A^2\rangle \;, \label{fourierff} \\
\langle r_A^2\rangle &=&   \int d^2r_A  r_A^2 \rho(r_A) \; , 
\end{eqnarray}
where we assumed a normalized one-body density and used that all odd
powers of $\vec{Q}$ vanish due to spherical symmetry.  The first order
derivative with respect to $Q^2$ then gives the desired mean-square
radius.  In the same way we obtain  
\begin{eqnarray}
 \langle r_{AB}^2\rangle  &=& \int d^2r_{AB}  r_{AB}^2 {\bar \rho}(r_{AB}) \; .
\end{eqnarray}
from the Fourier transform, $F_{AB}(Q^2)$ of ${\bar
  \rho}(\vec{r}_{AB})$.  These Fourier transforms are related to the
momentum-space wave function in Eq.(\ref{psiab}) by
\begin{eqnarray}
\label{fa}
\nonumber
F_A(Q^2)&=&\int d^2qd^2p\Psi_{AB,A}\left(\vec{q}+\frac{\vec{Q}}{2},\vec{p}\right)\\
&&\times\Psi_{AB,A}\left(\vec{q}-\frac{\vec{Q}}{2},\vec{p}\right),\\
\nonumber
F_{AB}(Q^2)&=&\int d^2qd^2p\Psi_{AB,A}\left(\vec{q},\vec{p}+\frac{\vec{Q}}{2}\right)\\
&&\times\Psi_{AB,A}\left(\vec{q},\vec{p}-\frac{\vec{Q}}{2}\right),
\label{fab}
\end{eqnarray}
which can be verified by directly inserting the definitions of
$\rho(r_A)$ and $\rho(r_{AB})$ from Eqs.(\ref{rhoa}) and (\ref{rhoab})
into Eq.(\ref{fourierff}) and the corresponding definition of
$F_{AB}(Q^2)$.

To calculate the remaining mean-square radii we express the three-body
wave function in the set of Jacobi momenta of the $B$-particle,
$\Psi_{AA,B}(\vec{q}_B,\vec{p}_B)\equiv
\langle\vec{q}_B,\vec{p}_B\,|\Psi\rangle$.  The mean-square radii,
$\langle r_B^2 \rangle$ and $\langle r_{AA}^2 \rangle$, are then
obtained as first order derivatives with respect to $Q^2$ of the
Fourier transforms, $F_B(Q^2)$ and $F_{AA}(Q^2)$, defined by replacing
$\Psi_{AB,A}$ with $\Psi_{AA,B}$ in Eqs.(\ref{fa}) and (\ref{fab}).

The procedure is now to solve the three-body equations in
Eqs.(\ref{chib}) and (\ref{chia}) and obtain three-body energy,
spectator functions, and the total wave function in Eq.(\ref{psiab}).
The next steps are to calculate numerically $F_A$, $F_B$, $F_{AA}$ and
$F_{AB}$ and the necessary first order derivatives of these functions
of the squared momenta.

\section{Mass dependence}
\label{mass}

The three-body wave function is completely determined from the two
ratios of particle masses, ${\cal A} = m_B/m_A$, and two-body
bound-state energies, $E_{AA}/E_{AB}$, both negative.  To extract
meaningful structure properties it is crucial to know all independent
distances in a given system, as illustrated in Fig.\ref{figradii}.  We
calculate energies and radii of the lowest-lying excited states as
functions of ${\cal A}$ for an energy ratio of $E_{AA}/E_{AB} = 1$.
Each of the two-body energies results from the related reduced mass,
which therefore is varied in our investigations.  The calculated
results therefore illustrate typical behavior of mean-square radii and
structure in general.  We confirm this expectation with test
calculations of varying energy ratio.

\subsection{Three-body energies}
\label{ener}

\begin{figure}[htb!]
\epsfig{file=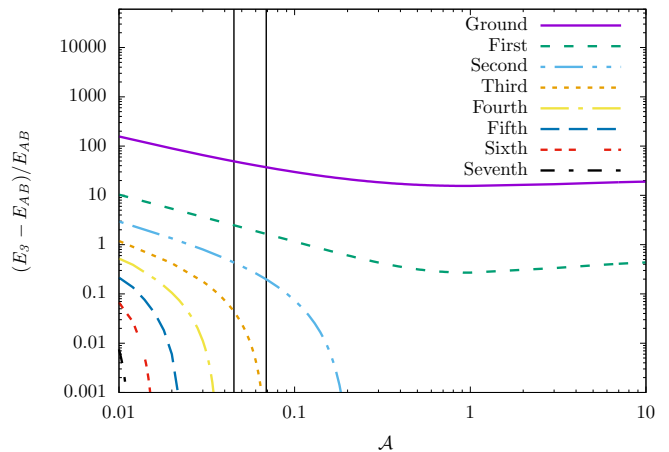,width=\columnwidth}
\caption{Low-energy spectrum of an $AAB$ system as a function of the
  mass ratio ${\cal A}=m_B/m_A$.  The two two-body energies are equal,
  $E_{AB}=E_{AA}$, and the three-body energy is $E_3$.  The
  energies on the $y$-axis are given relative to the two-body bound
  state energies, $E_{AB}$.  The vertical lines
  indicate the mass ratios ${\cal A}=6/133$ and ${\cal A}=6/87$
  corresponding to the systems $^6$Li-$^{133}$Cs-$^{133}$Cs and
  $^6$Li-$^{87}$Rb-$^{87}$Rb. }
\label{figenergy}
\end{figure}

The energy is for most quantum states the all-important characterizing
quantity which is necessary to understand first.  Universal structure
of weakly bound systems often relates spatial extension and binding
energy \cite{nie99}.  We therefore first focus on the energy
dependence of the system, $AAB$, sketched in Fig. \ref{figradii}.  We
solve the coupled set of integral equations, Eqs.(\ref{chib}) and
(\ref{chia}), and show the calculated energy spectrum in
Fig. \ref{figenergy}.  For ${\cal A}=1$ we recover the well known
results that only ground and first excited states are bound with
universally given energies, $E_3=16.52E_{AB}$ and $E_3=1.27E_{AB}$.
These energies are predicted by a number of entirely different
calculations \cite{bruchpra1979,nie99,bellottipra2012}.

Increasing the mass ratio above unity, ${\cal A} \geq 1$, the two
bound states remain with moderate relative energy variations.  This
may be understood from the limit of one very heavy particle surrounded
by two light masses moving around the center-of-mass in a roughly
mass-independent field.  Superficially this resembles a helium atom
with two electrons moving around the four bound nucleons.  However,
the interactions differ enormously from the long-range Coulomb
potential to the zero-range interaction.  The properties in 2D also
differ enormously from the 3D calculations of a helium atom.  These
differences are evident from the order-of-magnitude-larger ground-state 
binding of three particles compared to two particles.

In the other limit of small ${\cal A}$ we find a tremendous increase
of bound states with decreasing mass ratio ${\cal A}$.  The smallest
mass ratio we consider is ${\cal A}=0.01$, where the system displays
eight bound states.  The excited three-body states disappear into the
continuum of positive energies as ${\cal A}$ increases from very
small values.  The rather dramatic mass dependence for small ${\cal
  A}$ can be understood as a consequence of the increase of the
effective interaction generated by the light particle.  This has been
known since 1979 \cite{fonsecaNPA78} and 1980 \cite{limZPA80} for
systems in three and two dimensions, respectively.

Recently, a different technique was used to derive an effective
two-body potential in two dimensions \cite{bellottijpb2013}.  This
derivation assumed the Born-Oppenheimer approximation whose validity
only is justified for ${\cal A} \ll 1$.  After integrating out the
coordinate of particle $B$, an effective hamiltonian is left for the
relative motion of the two heavy $A$-particles, that is
\begin{equation} \label{HBO}
 H_{BO} =  - \frac{\hbar^2}{2\mu_{AA}} \Delta_{\vec{r}_{AA}} + V_{AA}(r_{AA}) + 
 V^{(BO)}(r_{AA}) \; ,
\end{equation}
where $\vec{r}_{AA}$ is the relative coordinate between the two
$A$-particles, $\mu_{AA}$ and $V_{AA}(r)$ are their reduced mass and
two-body potential, and $V^{(BO)}(r_{AA})$ is the strongly
${\cal A}$-depending Born-Oppenheimer potential resulting from the
light third particle, $B$.  

The Born-Oppenheimer potential is obtained by standard derivation in
three-body physics, that is with two centers at a fixed distance
interacting with the light particle through zero-range interactions.
This quantum mechanical problem in 2D is solved in the center-of-mass
of the total system.  The result is in Ref. \cite{bellottijpb2013}
found to be
\begin{equation}
\ln \frac{\left\vert V^{(BO)}(R)\right\vert }{\left\vert E_{AB} \right\vert} =2 K_0 \left(\sqrt{\frac{2\mu_{AA,B}|V^{(BO)}(R)|}{\hbar^2}} r_{AA} \right)  \ , \label{eq.51}
\end{equation}
where $K_0$ is a Bessel function, $\mu_{AA,B}$ is the reduced mass of
$B$ versus the $AA$-system, and $R = r_{AA} \sqrt{m_A
  |E_{AB}|/\hbar^2}$ with $\mu_{AA,B}$.  This analytical result was in
Ref. \cite{bellottijpb2013} shown to be well approximated as a function
of ${\cal A}=m_B/m_A$ by simpler expressions in both limits of small
and large distances.  The ${\cal A}$-dependence is through the
combination
\begin{equation} \label{BOmass}
 m_{ef} = \bigg(\frac{4{\cal A}}{2+{\cal A}} \bigg)^{1/2} \; ,
\end{equation}
and the results from Ref.\cite{bellottijpb2013} are
\begin{equation} \label{BOpota}
V^{(BO)}(r_{AA}) \approx - \frac{2 |E_{AB}|\exp(-\gamma) }{Rm_{ef}} \; 
\end{equation}
for $R m_{ef} \leq 1.15 $, where $\gamma = 0.5772156649$ is Euler's
constant, and
\begin{equation}
V^{(BO)}(r_{AA}) \approx  - |E_{AB}| \Big( 1 + 
\frac{\sqrt{2\pi}\exp(-R m_{ef})}{\sqrt{R m_{ef}}}\Big)  
\label{BOpotb}
\end{equation}
for $R m_{ef} \geq 1.15 $.  These approximations are accurate to
better than $10\%$, where the largest deviations found around $R
m_{ef} = 1.15 $ \cite{bellottijpb2013}.

The short-distance part exhibits an attractive Coulombic behavior in
Eq. (\ref{BOpota}).  The large-distance behavior in Eq. (\ref{BOpotb})
is exponentially convergent towards the negative two-body energy,
$E_{AB}$, with a length scale proportional to $1/m_{ef}$.  Thus, only
a finite number of bound states below $E_{AB}$ are possible, although
that number would increase without limit since the $1/m_{ef}$ diverges
with vanishing mass ratio ${\cal A}$.  However, we are mostly
interested in the lowest energy states which necessarily are located
in the strongly attractive region at smaller distances.  The
small-distance Coulomb-like behavior is controlled by an effective
charge squared, $Z_{ef}^2$, defined by
\begin{equation} \label{charge2}
  Z_{ef}^2 = 
 \frac{ 2 \hbar \exp(-\gamma) \sqrt{|E_{AB}|}}{m_{ef} \sqrt{m_A}} \; .
\end{equation}
The kinetic energy operator in Eq. (\ref{HBO}) has a mass of $\mu_{AA}
= m_A/2 $, and in two dimensions the crucial negative centrifugal
barrier term corresponds to the angular momentum quantum number of
$\ell = -1/2$.  The Coulomb energy spectrum for this small-distance
behavior is therefore
\begin{eqnarray} \label{energy}
\nonumber
E^{(BO)}_3 &=& - \frac{Z_{ef}^4 \mu_{AA}}{2\hbar^2} \frac{1}{(n_r+\ell+1)^2}\\ 
\nonumber
  &=&  \frac{(1+{\cal A}/2)}{2{\cal A}}\frac{\exp(-2\gamma)|E_{AB}|}{(n_r+\ell+1)^2}\\
 &=&  \frac{0.630473504 (1+{\cal A}/2) |E_{AB}| }{{\cal A}(2n_r+1)^2}  \; .
\end{eqnarray}
The entire Coulomb spectrum is then obtained with $n_r =0,1,2,...$ and
$\ell=-1/2,1/2,3/2,5/2...$

\begin{table}
\begin{tabular}{@{}|c|c|c|c|c|}
\hline
& ${\cal A}$ & $E^{(BO)}_{3}/E_{AB}$ & $E^{(NI)}_{3}/E_{AB}$ & $E_{3}/E_{AB}$ \\
\hline
Ground & 0.01       & 63.04          & 53.07         & 157.56  \\
       & 0.02       & 31.52          & 27.76         & 90.75   \\ \hline
First  & 0.01       & 7.00           & 7.72          & 11.34   \\
       & 0.02       & 3.50           & 4.25          & 6.36    \\ \hline
Second & 0.01       & 2.52           & 3.21          & 3.94    \\
       & 0.02       & 1.26           & 1.92          & 2.34    \\ \hline
Third  & 0.01       & 1.28           & 1.92          & 2.18    \\
       & 0.02       & 0.64           & 1.28          & 1.42    \\
\hline
\end{tabular}
\caption{\label{I}Energies of ground and first excited states for the mass
  ratios ${\cal A} =$ 0.01, 0.02 for the analytic Born-Oppenheimer
  approximation, $E^{(BO)}_{3}/E_{AB}$ (in Eq. (\ref{energy})), the numerical
  results both without, $E^{(NI)}_{3}/E_{AB}$, and with, $E_{3}/E_{AB}$, interaction
  between the two heavy $A$-particles.}
\end{table}

We can use Eq. (\ref{energy}) as a useful reference spectrum for
comparison to the energies in the limit of ${\cal A} \ll 1$, shown in
Fig. \ref{figenergy}.  To do this we have to distinguish between the
present numerical calculations and the pure Coulomb spectrum obtained
by neglecting the short-range potential, $V_{AA}$, between the two
heavy $A$-particles.  We compare these results in the first two
columns of Table \ref{I}.  The pure Born-Oppenheimer Coulomb estimate
of the ground state binding energy is $15$~\% larger than the
numerically calculated value.  On the other hand, the Coulomb
estimates give too small binding for all the excited states.  They
exploit distances outside the Coulomb-like attractive region.

The ground state deviations were also found in
Ref. \cite{bellottijpb2013}.  They are necessarily due to the
approximations in the Born-Oppenheimer procedure, since the accuracy
of our numerical computations are fractions of permille.  The most
obvious reason for these differences is the neglect of the
non-vanishing term arising from the heavy-heavy kinetic energy
operator.  The Born-Oppenheimer wave function depends first of all on
the fast coordinate but there is still also a dependence on the slowly
varying coordinate, $r_{AA}$.  This term is not included in our
Born-Oppenheimer calculation, where the inherent two-step quantization
procedure in any case is an approximation.

It is illuminating to compare three-body calculations with the
hyperspherical adiabatic expansion method, where the hyperradial part
of the kinetic energy operator gives rise to an analogous term
\cite{nielsenpr2001}.  Inclusion or not of this term, in the adiabatic
expansion with only one adiabatic potential, provides an upper or
lower bound on the correct energy \cite{nielsenpr2001}.  Including
more and more of the higher-lying potentials in this method provides a
fully converged solution.  The Born-Oppenheimer potential is analogous
to the one-potential approximation, which overbinds in agreement with
the bounds of the mentioned hyperspherical adiabatic method.  The
simplicity of the Born-Oppenheimer results allows qualitative
understanding of spectra and structure, which is appealing even though
the Faddeev calculations could speak for themselves.  In any case the
present Born-Oppenheimer approximation suffice for our purpose.

We now include the $AA$-interaction and focus on the results in
Fig. \ref{figenergy} where each level for small ${\cal A}$ roughly
follows the predictions in Eq. (\ref{energy}).  However, the
calculated ground state binding energies for both values, ${\cal A} =
0.01, 0.02$, are much larger than the pure Coulomb spectrum, see Table
\ref{I}.  The strong short-range $AA$-attraction has its largest effect on
the lowest-lying levels, that is more binding by factors of about $3$
and $1.5$ for ground and first excited states, respectively.  The
higher-lying excited states shown in Table \ref{I} and
Fig. \ref{figenergy} feel the long-range non-Coulombic
Born-Oppenheimer potential and exhibit a less systematic behavior.  As
${\cal A}$ increases, the bound states move up in energy, the number
decreases, and the Coulomb potential supports fewer and fewer bound
states.

\subsection{Sizes }

\begin{figure}[htb!]
\epsfig{file=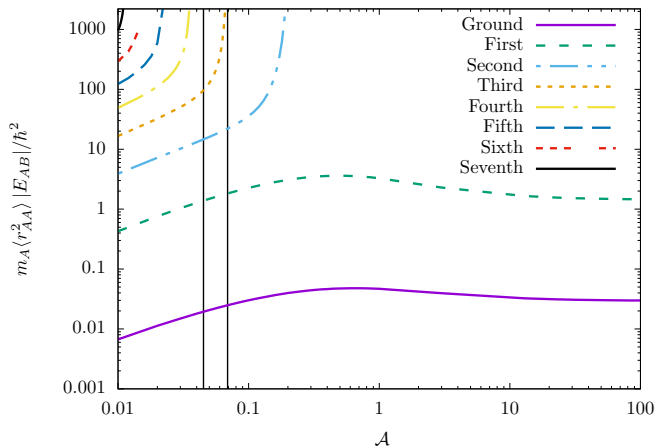,width=\columnwidth}
\caption{Dimensionless product $m_A\langle r_{AA}^2\rangle
  |E_{AB}|/\hbar^2$ ($E_{AA} = E_{AB}$) as a function of the mass
  ratio ${\cal A}$. As ${\cal A}$ is increased the radii diverge at
  the threshold where the excited states disappear. The remaining
  ground and first excited states assume a constant value as ${\cal
    A}\rightarrow\infty$.  Vertical lines are the mass ratios
  corresponding to the systems $^6$Li-$^{133}$Cs-$^{133}$Cs and
  $^6$Li-$^{87}$Rb-$^{87}$Rb.}
\label{figraa}
\end{figure}

Equipped with an understanding of the energy spectrum as function of
the mass ratio we turn to the corresponding underlying structure.  We
show first in Fig. \ref{figraa} the calculated mean-square radial
distances, $<r^2_{AA}>$, between the two identical particles.  We
notice an almost reflected behavior compared to the three-body
energies.  The radii decrease while the binding energies increase.
The two lowest bound states are present for all mass ratios, whereas
the sizes for the higher-lying excited states diverge as their energies
reach their thresholds for binding.

These radii are for small mass ratios, ${\cal A} \ll 1$, related to the
Coulombic orbits discussed in connection with Fig. \ref{figenergy}.
The Coulomb radii are given analytically by
\begin{eqnarray} 
\nonumber
 &&<r^2_{AA}> =  \frac{1}{2} \bigg(\frac{\hbar^2}{\mu_{AA} Z_{ef}^2 }\bigg)^2 
 (n_r+\ell+1)^2\\ 
 &&\;\;\;\;\;\;\;\;\;\;\;\;\;\;\;\;\;\times\bigg(5(n_r+\ell+1)^2 +1 -3\ell(\ell +1)\bigg)  
  \nonumber \\ \label{radi}
 &&=\frac{3.1722\hbar^2{\cal A}(2n_r+1)^2 (5(2n_r+1)^2 + 7)}{16 m_A |E_{AB}|(1+{\cal A}/2)}
\end{eqnarray}
where we used Eq.(\ref{charge2}) and $\ell = -1/2$.


\begin{table}
\begin{tabular}{@{}|c|c|c|c|c|}
\hline
& ${\cal A}$ & $\langle\, r^{(BO)2}_{AA}\,\rangle/R_u^2$ & $\langle\, r^{(NI)2}_{AA}\,\rangle/R_u^2$ & $\langle r^{2}_{AA}\rangle/R_u^2$ \\
\hline
Ground & 0.01       & 0.023          & 0.029         & 0.006  \\
       & 0.02       & 0.047          & 0.056         & 0.011   \\ \hline
First  & 0.01       & 0.923           & 0.916          & 0.427   \\
       & 0.02       & 1.837           & 1.767          & 0.757    \\ \hline
Second & 0.01       & 6.510           & 6.268          & 3.911    \\
       & 0.02       & 12.955           & 12.254          & 7.201    \\ \hline
Third  & 0.01       & 24.359           & 23.381          & 16.594    \\
       &0.02        &48.478        & 48.534          & 32.219    \\
\hline
\end{tabular}
\caption{\label{II} Mean-square-radii in the Born-Oppenheimer
  approximation, $\langle r^{(BO)2}_{AA}\rangle$, from
  Eq. (\ref{radi}), and numerical results repectively with and without
  the interaction between the two $A$-particles, $\langle
  r^{2}_{AA}\rangle$ and $\langle r^{(NI)2}_{AA}\rangle$.  All these
  lengths are in units of $R_u \equiv \hbar/\sqrt{m_A |E_{AB}|}$.}
\end{table}

Let us first consider the radii of the states obtained for the
smallest mass ratios of ${\cal A}=$ 0.01, 0.02.  We compare with the
Born-Oppenheimer estimates precisely as we did for the energies in
Table \ref{II}.  As usual for universal structures the radii have the
opposite behavior to the energies, that is the mean square radii are
largest for the smallest binding energies.  The Born-Oppenheimer
results for the ground state are about $15$~\% smaller than obtained
from the full numerical calculation without an $AA$ interaction.
These deviations are again probably mostly due to the neglect of the
heavy-heavy kinetic energy operator on the light particle
Born-Oppenheimer wave function.  For the excited states the same
comparison show larger Born-Oppenheimer radii.  The observed opposite
tendencies of energy and radii are reflected in the formula in
Eqs.~(\ref{energy}) and (\ref{radi}) where the product is state
independent apart from the last factor in Eq. (\ref{radi}).

We now turn to the full calculation with sizable two-body energies
between all three pairs of particles, shown in the last column of
Table \ref{II}.  The ground state mean-square radius is a factor of
about $3.5$ smaller than derived from the Coulomb estimate in
Eq. (\ref{radi}).  This deviation is again consistent with the similar
larger binding energy obtained for that state with the same
interaction.  The first excited state is only smaller than the Coulomb
estimate by a factor of $2.2$.  The following two higher-lying states
have radii rather similar to the Coulomb estimates.  These states
extend beyond the Coulomb region and into the region of the more
confining large-distance Born-Oppenheimer potential from
Eq. (\ref{BOpotb}).  The effect is a comparably smaller spatial
extension which is very close to the results from Eq. (\ref{radi}).

\begin{figure}[htb!]
\epsfig{file=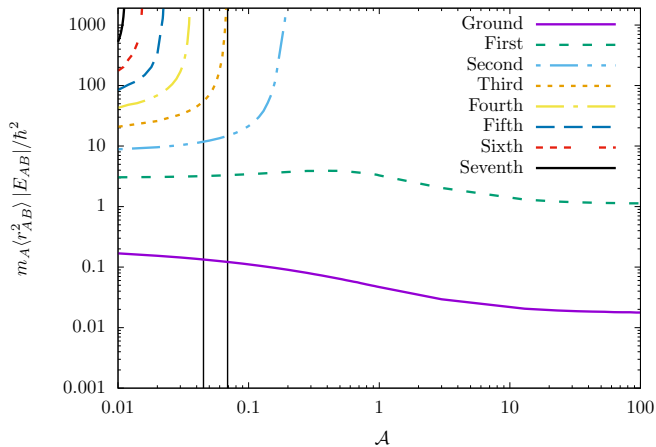,width=\columnwidth}
\caption{The same as Fig. \ref{figraa} when $\langle r_{AB}^2\rangle$
  replaces $\langle r_{AA}^2\rangle$. } 
\label{figrab}
\end{figure}

At least one more distance is required to characterize the geometric
structure of the three-body system.  We first choose the distance,
$r_{AB}$, between unequal pairs of particles, $A$ and $B$.  The
mean-square radius is a quantum mechanical expectation value where the
two identical particles cannot be distinguished.  The results shown in
Fig. \ref{figrab} are therefore averages over distances between
particle $B$ and the two $A$-particles.  We notice first that the $AB$
mean-square distance is a flat or slightly increasing function with
decreasing small mass ratios for both ground and first excited states.
The higher-lying states show the opposite tendency of marginal
decrease.  Furthermore the $r_{AB}$-value in Fig. \ref{figrab} is much
larger than the $r_{AA}$ radius in Fig. \ref{figraa}, although with a
difference decreasing with excitation energy.  To understand this we
cannot turn to the Born-Oppenheimer calculations which only provides
information about the $AA$ system while the $B$-coordinate is
integrated out.  In the limit of small mass ratios the light
$B$-particle is very little concerned with the slow relative $AA$
motion.  The $B$-particle moves much faster and almost independently
in an orbit of much larger radius.

The behavior changes drastically with increasing mass ratios.
First the radii of the excited states diverge at their thresholds of
binding.  Second, the distance $r_{AB}$ decreases with increasing
${\cal A}$ for the two bound states.  The results for ${\cal A} = 1$
agree with the known mean-square ratio of about $70$ between the two
bound states \cite{nielsenpr2001}.  As ${\cal A}$ increases above $1$
we see that the radii only changes very little precisely consistent
with the same behavior as shown by the binding energies.

\subsection{Structure}

The sizes reflect the structures described by the corresponding wave
functions.  We would like, schematically and optimistically, to
reproduce the relative as well as the absolute sizes in
figs.~\ref{figraa} and \ref{figrab} by interpretation with appropriate
wave functions and/or geometric configurations.  We focus on a large
mass ratio of ${\cal A} = 100$ where the ratio between mean-square
distances of $AA$ and $AB$-particles is $1.67$ and $1.3$ for ground
and excited states, respectively.  A fully symmetric wave function
in hyperspherical coordinates corresponds to
$\rho^{-3/2} \exp(-\kappa \rho)$ where $\rho$ in the present case is
defined by
\begin{eqnarray} \label{rho}
\nonumber
(2m_A +m_B) \rho^2 &=&   m_A (\vec{r}_{A}- \vec{r}_{A'})^2
  +  m_B (\vec{r}_{A}- \vec{r}_{B})^2\\ 
  &&+  m_B (\vec{r}_{A'}- \vec{r}_{B})^2 \; ,
\end{eqnarray}
where the $\kappa$-value is given by $\hbar^2 \kappa^2 = 2 m_A|E_3|$.
This wave function has the correct large-distance asymptotic behavior.
The mean-square radius between a pair of particles, $i$ and $k$, is
then from this very simplified symmetric wave function found by
straightforward calculations to give
\begin{equation} \label{radsym}
  \langle r_{ik}^2\rangle = \hbar^2/(8\mu_{ik}|E_3|) \; ,
\end{equation}
where $\mu_{ik}$ is the reduced mass of particles $i$ and $k$
($i,k$=A,A',B).  We shall refer to it as a symmetric estimate as it is
obtained from the fully symmetric wave function.  This schematic
estimate from Eq. (\ref{radsym}) for the $AA$ mean-square radii of
ground and excited states is smaller by a factor of $2.6$ and $8.8$,
respectively.  The corresponding ratios for the $AB$ mean-square radii
are $3.1$ and $14$ which first of all reflect the different reduced
mass dependence.  These numbers are for squared radii, and the linear
distances, obtained by taking the square root, then only deviate by
factors varying from $1.5$ to about $4$.

Extreme asymmetric structures may be viewed with two particles in a bound
state at small distance and the third particle more loosely bound at a
larger distance.  To produce a finite $AA$-distance the configuration
must correspond to a structure like $A-(AB)$ (see Eq. \ref{radi}).  
The mean-square radius is then $\hbar^2/(4\mu_{A,AB}|E_{A,AB}|)$ which 
again is inversely proportional to the two-body reduced mass of the distant 
$A$-particle relative to the stronger bound $AB$-entity.  However, the absolute
value of this estimate lies between those of ground and excited states
for a reasonable choice of $E_{A,AB} \approx 2 \times E_{AB}$.

Both extreme structures of total symmetry and extreme asymmetry fail
to reproduce the calculated moments.  Furthermore, the large
difference between ground and excited states indicate a sizable
structure variation.  The ground state is within a factor of two from
the symmetric estimate, whereas the excited state extends in size
substantially beyond the inverse three-body energy relation in
Eq. (\ref{radsym}).  The latter is more reminiscent of the huge
increase of radii found in the Efimov states which in three dimensions
are coherent superpositions of asymmetric geometric structure.

\begin{figure}[htb!]
\epsfig{file=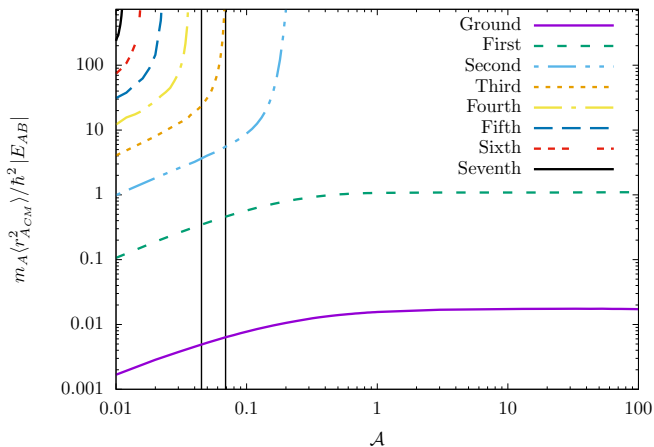,width=\columnwidth}
\caption{Same as Fig. \ref{figraa} when the mean-square distance,
  $\langle r_{A_{CM}}^2\rangle$, from C.M to particle $A$ replaces
  $\langle r_{AA}^2\rangle$, as ${\cal A}\rightarrow\infty$ $\langle
  r_{A_{CM}}^2\rangle\rightarrow\langle r_{AB}^2\rangle$.} 
\label{figracm}
\end{figure}

Let us finally consider asymmetric structures where a finite average
radius is assumed between the strongest bound $AB$ two-body
substructure.  We first assume a given distance, $r_{AA'}$, between
the two $A$-particles, where the $B$-particle moves in a uniform
circular orbit of size $r_{A'B}$
around one of the $A'$-particles. Then we estimate the
mean-square distance $\langle r_{AB}^2 \rangle \approx (\langle
r_{AA'}^2 \rangle + \langle r_{A'B}^2 \rangle)/2$, where $1/2$ is from
adding the two symmetrized configurations. To reproduce the calculated
mean-square radii we need $\langle r_{AB}^2 \rangle / \langle r_{AA}^2
\rangle \approx 0.2, 0.5$ for ground and excited states.  These values
show a strong tendency towards symmetric configurations.

The sizes are often measured as distances relative to the total
center-of-mass. This only provides little additional information but
it is very illustrative besides serving as a consistency check on our
understanding of the underlying structures. We show in
Fig. \ref{figracm} the mean-square radial distance of particles $A$
from the center-of-mass.  The two limits of ${\cal A}$ differ very
much from each other.  Varying ${\cal A}$ from small to large values
causes the three-body center-of-mass to move from the center-of-mass of the
$AA$ system to the center of particle $B$.  This implies by geometric
reasoning that $\sqrt{\langle r_{A}^2 \rangle}$ approaches 
$\sqrt{\langle r_{AA}^2\rangle}/2$ and $\sqrt{\langle r_{AB}^2\rangle}$ 
in these two limits, respectively.  These predictions are confirmed by comparing
Fig. \ref{figracm} with the results in Figs. \ref{figraa} and
\ref{figrab}, that is observing that $\langle r_{A}^2\rangle \rightarrow
\langle r_{AA}^2\rangle/4$ and $\langle r_{A}^2\rangle \rightarrow \langle r_{AB}^2\rangle$ 
for small and large ${\cal A}$.

\begin{figure}[htb!]
\epsfig{file=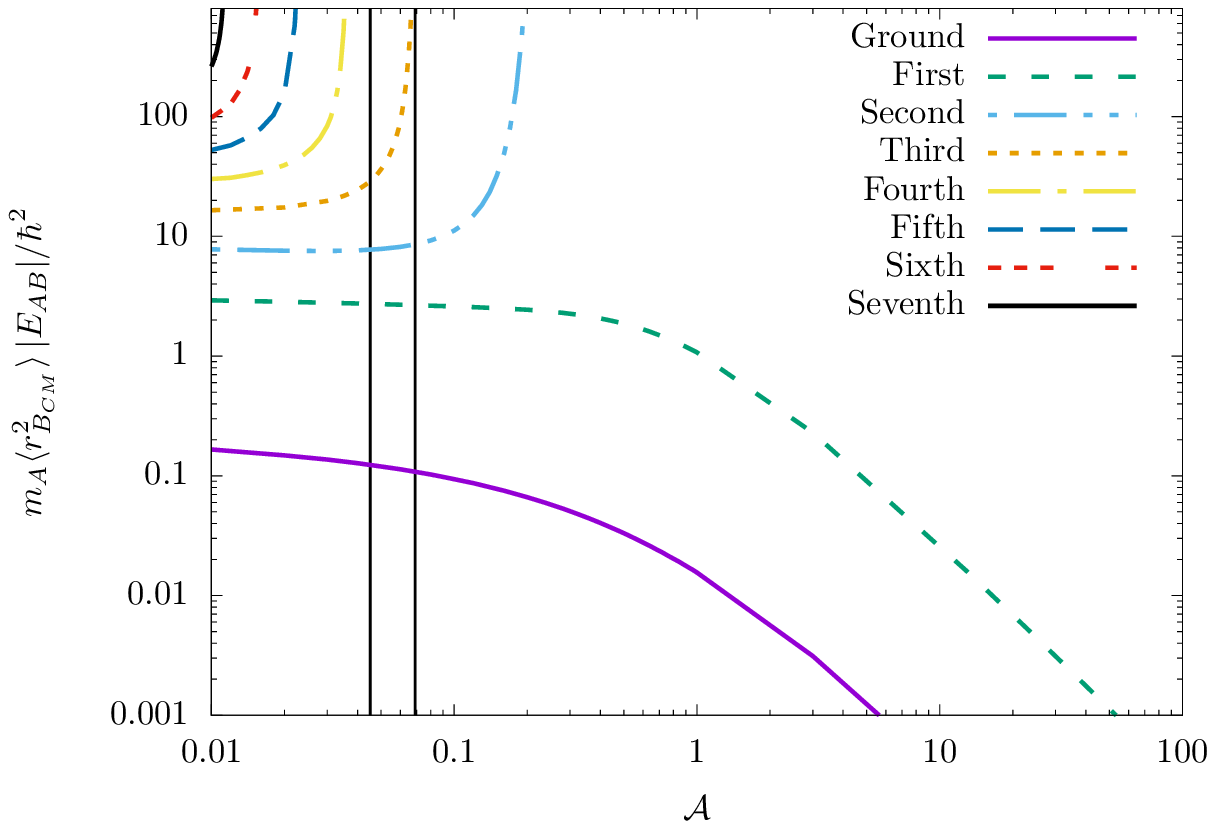,width=\columnwidth}
\caption{Same as Fig. \ref{figracm}, but when the mean-square distance,
  $\langle r_{B_{CM}}^2\rangle$, from C.M to particle $B$ replaces
  $\langle r_{A_{CM}}^2\rangle$, as ${\cal A}\rightarrow\infty$
  $\langle r_{B_{CM}}^2\rangle\rightarrow0$.} 
\label{figrbcm}
\end{figure}

The different behavior of $\langle r_{B}^2 \rangle$ is seen in
Fig. \ref{figrbcm}.  The geometry now predicts that $r_{B}$ should
vanish for large ${\cal A}$ since the total center-of-mass coincides
with the center of particle $B$.  In the other limit of small ${\cal
  A}$ we know that $\langle r_{AA}^2 \rangle \ll \langle r_{AB}^2
\rangle$ and consequently $\langle r_{B}^2 \rangle$ should approach
$\langle r_{AB}^2 \rangle$, since the $AA$ system looks like an entity from
far away where particle $B$ is located.  Again these predictions are
confirmed numerically by comparing Figs. \ref{figrab} and
\ref{figrbcm}.

There is a lack of experimental information concerning the sizes of
these systems in two dimensions. However, we are able to compare our
results for ${\cal A}=1$ with those from Ref. \cite{nielsenpr2001} as
follows. The spatial extension of a system can be measured by radial
moments. Often the second moment, the root-mean-square radius, is used
as an average measure.  Different distributions can be of interest as
exemplified by average charge- or mass-radii.  For a self-supported
system of bound particles the center-of-mass is conserved and it is
natural to measure all distances with respect to this point,
$\vec{R}_{CM}$.  The corresponding expectation value is related to the
root-mean-square radius, $R_{rms}$, of the mass distribution as
defined for our three-body system by
\begin{eqnarray}
\nonumber
 &&R_{rms}^2 \sum_{k=A,A',B} m_k = \sum_{k=A,A',B} m_k \langle (\vec{r}_{k}  - \vec{R}_{CM})^2\rangle \\   \label{rms} 
 &&=\sum_{k=A,A',B} m_k \langle \vec{r}^2_{k_{CM}}\rangle  \; ,
\end{eqnarray}
where the vectors, $\vec{r}_{k_{CM}}$, by definition connects particle
$k$ and the center-of-mass of the three-body system.  The three-body
center-of-mass lies on the line connecting particle $A$ with the
center-of-mass of the remaining pair of $AB$-particles, and
analogousluy for particle $B$.  We therefore have the relations
\begin{eqnarray}
\langle r_{A_{CM}}^2 \rangle = \left(\frac{1+{\cal A}}{2+{\cal A}}\right)^2 \langle r_{A}^2 \rangle \;\; ,\\
\label{racm}  \;\; \langle  r_{B_{CM}}^2 \rangle = \left(\frac{2}{2+{\cal A}}\right)^2 \langle r_B^2\rangle \;,
\end{eqnarray}
which can be used to express $R_{rms}^2$ from Eq. (\ref{rms}) in terms
of different mean-square radii, that is
\begin{eqnarray}
\nonumber
 R_{rms}^2 &=& \frac{2 \langle r_{A_{CM}}^2 \rangle  + 
 {\cal A}  \langle  r_{B_{CM}}^2 \rangle  } {2 + {\cal A}}\\  
 &=& 2 \frac{(1 + {\cal A})^2 }{(2 + {\cal A})^3} \langle r_{A}^2 \rangle +
 \frac{ 4 {\cal A} }{(2 + {\cal A})^3} \langle r_{B}^2 \rangle \;.
\end{eqnarray}

For three identical particles, where ${\cal A} = 1$ and $\langle
r_{A}^2 \rangle = \langle r_{B}^2 \rangle$, this relation reduces to
\begin{eqnarray}
R_{rms} &=&\frac{2}{3} \sqrt{ \langle r_{A}^2 \rangle } \;,
\end{eqnarray}
with $\langle r_{A}^2 \rangle$ plotted in Fig. \ref{figracm}.

The two bound three-body states labeled $0$ and $1$ are in
\cite{nielsenpr2001} found to have root-mean-square radii given by
\begin{eqnarray}
 R_{rms}^{(0)} = 0.111 a = 0.125  \sqrt{\frac{\hbar^2}{m_A |E_{AB}|}}
\; \; , \; \;\\
 R_{rms}^{(1)} = 0.927 a  = 1.041  \sqrt{\frac{\hbar^2}{m_A |E_{AB}|}} \, ,
\end{eqnarray}
where the two-dimensional scattering length, $a$, is related to the
two-body energy $E_{AA} = E_{AB} = -4 e^{-2 \gamma} \hbar^2 /(m_A
a^2)$. These values match exactly our numerical results.

\subsection{Threshold behavior}

\begin{figure}[htb!]
\epsfig{file=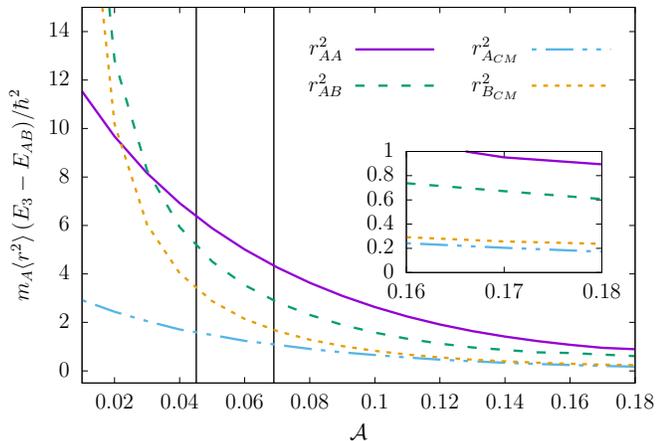,width=\columnwidth}
\caption{Threshold behavior of different mean-square radii for the
  second excited state as a function of mass ratio.  The divergent
  mean-square radii multiplied by $E_3-E_{AB}$ reach constant values
  at threshold.  Vertical lines are the mass ratios corresponding to
  the systems $^6$Li-$^{133}$Cs-$^{133}$Cs and
  $^6$Li-$^{87}$Rb-$^{87}$Rb.}
\label{figdiv}
\end{figure}

We want now to investigate in more detail how the structure varies
with mass ratio when the energy of a state approaches its threshold
for binding.  We know that in three dimensions the sizes of bosonic
three-body systems diverge logarithmically at their three-body energy
thresholds. In contrast, the sizes remain finite for brunnian systems 
(all pairs are unbound) with more than three bosons \cite{brunnian}, 
even at their corresponding thresholds of binding.  In these cases, 
the thresholds are reached from the bound side by decreasing the 
two-body attraction.  In the present study we
are decreasing the mass ratio, and in practice decreasing the
effective interaction between the constituents of the system.

Again we use the mean-square radii as indicative measures.  We focus
on the second excited state which is the lowest-lying state with
diverging size.  Its behavior is repeated by the other excited
states, which we therefore do not need to discuss in this paper.  We
show in Fig. \ref{figdiv} the results as functions of ${\cal A}$ for
the different radii of the second excited state.  We multiply the
radii by the energy deviation from the two-body threshold,
$|E_3-E_{AB}|$, in order to extract the behavior of the divergence.  The
striking result is that this product converges towards a constant as
the threshold is approached. Thus, the divergence of the mean-square
radius is inversely proportional to $E_3-E_{AB}$.  This behavior is
well-known for weakly bound two-body halos in three dimensions
\cite{jensenrmp2004}, and shown in Ref. \cite{nielsenpr2001} to be
valid also in two dimensions.  The implication is that the threshold
structure must correspond to one particle moving away from a bound
and spatially confined subsystem, which in our case only can be a bound
dimer.

Two structures are possible at threshold, that is, either particle $A$
or particle $B$ is ejected while the remaining pair settles in their
ground state.  In both cases the threshold structure resembles
two-body systems, that is, $(AA)-B$ or $(AB)-A$ with corresponding
reduced masses $\mu_{AA,B} \approx 0.18 $ or $\mu_{AB,A} \approx
0.54$, where the numbers are obtained with a threshold value of
${\cal A} \approx 0.2$.

These structures would according to Ref. \cite{nielsenpr2001} lead to
the two-body divergence, $<r^2_{ik}> = \hbar^2/(3\mu_{ik}|E_{th}|)$,
where $\mu_{ik}$ and $|E_{th}|$ are two-body reduced mass and the
vanishing threshold energy, respectively.  This means that the curves
in Fig. \ref{figdiv} for $m_A<r^2_{ik}> |E_{th}|/\hbar^2 =
m_A/(3\mu_{ik})$ can be either $1.85$ or $0.62$.  Thus, $(m_A
<r^2_{AA}> |E_{th}|/\hbar^2, m_A<r^2_{AB}> |E_{th}|/\hbar^2)
\rightarrow (0.0,1.85)$ or $(m_A<r^2_{AA}> |E_{th}|/\hbar^2,
m_A<r^2_{AB}> |E_{th}|/\hbar^2) \rightarrow (0.62,0.62)$ are the two
corresponding two-body threshold structures, where either $B$ or $A$
are ejected.  The two identical finite values in the latter case are
due to the same distance between the ejected $A$-particle and both the
remaining $A$ and $B$ particles in a bound state.

Obviously the best match to the results displayed in
Fig. \ref{figdiv} is ejection of one of the $A$-particles.  In the
calculated expectation value is contained an average over the
distances between the two $A$-particles and particle $B$.  One of
these distances remains finite and does not contribute except through a
reduction of the probability by a factor of $2$.  This accounts for the
value of $\frac{m_A<r^2_{AB}>}{\hbar^2}(E_{3}-E_{AB})=0.6$, being only about two thirds 
of $\frac{m_A<r^2_{AA}>}{\hbar^2}(E_{3}-E_{AB})=0.9$, at threshold.

\section{Summary, Conclusions and Perspectives}
\label{conclusions}

We studied the structures of three-body systems in two spatial
dimensions. We concentrate on asymmetric systems formed by two
identical bosons and a third particle.  We assume spin-independent
interactions and all results are therefore also valid for two
identical fermions with spatially symmetric wave functions.  We use
two-body zero-range interactions and the Faddeev equations in momentum
space to solve the three-body problem.  Universal properties are then
emphasized due to the vanishing interaction range.  The three-body
system is characterized by two masses and two zero-range strength
parameters chosen to reproduce specified two-body energies.  The
equations only depend on two parameters, that is mass and energy
ratios.  In this paper we essentially only investigate the mass
dependence for equal two-body energies.

Our aim is to obtain information about universal structures.  We
calculate second order radial moments corresponding to two independent
distances within the three-body system.  Correlations between these
quantities are very indicative of the underlying dominating structure.
Relatively low orders of radial moments are substantially easier to
compute than the density distributions, especially when the starting
point is in momentum space.  We therefore first derive suitable
expressions for the desired radial moments expressed in terms of
single first-order derivatives of integrals over momentum-space wave
functions.

We first calculate and analyze the energy spectrum as the primary
characterizing quantity of any quantum system.  We keep equal two-body
energies. The number of bound states are always finite but varies from
2 and upwards, diverging as the relative mass of the distinguishable
particle decreases from very large to approaching zero.  Only
precisely two bound states are present as for equal masses until the
mass ratio has decreased to about $0.2$.  Then a third state appears
followed by numerous other states as the mass ratio decreases towards
zero.  This agrees qualitatively with similar previous observations,
as well as quantitatively for the known case of equal masses.

We analyze the spectra for small mass ratio by use of the
Born-Oppenheimer approximation of the potential.  The effective
interaction generated by the light particle has Coulomb behavior for
small distances between particles of the heavier pair.  Combining with
the kinetic energy operator in two dimensions the Coulomb energy
spectrum appears for an angular momentum quantum number equal to
$-1/2$.  The small-distance region is the most attractive region,
where the short-range interaction between the two heavy particles also
is strongest.  Consequently this is where the lowest states are
located.

Considering this Coulomb potential calculated for small mass ratios
from the Born-Oppenheimer approximation, the three-body energies and
heavy-heavy mean-square radii are derived and compared to the full
calculations.  We see that the numerically obtained ground state is
more bound with a smaller radius than arising from the Coulomb
estimate.  This is due to the additional attractive heavy-heavy
zero-range interaction.  The first few of the following higher-lying
states still feel the heavy-heavy short-range interaction but they
have energies and radii closer to the Coulomb estimates. This reflects
that the effect of the zero-range attraction quickly decreases as the
states move to larger distances.  The higher-lying excited states
extend spatially beyond the Coulomb-like potential, but all bound
states are eventually confined within the exponentially decaying
Born-Oppenheimer potential at large distance.

The mean-square radii are to a large degree a reflection of the
energies, that is, small energy corresponds to a large radius and vice
versa.  This is the general observation for universal properties
when all angular momenta are zero or vanishingly small.  By comparing
two independent lengths in the same three-body system we can generate
a geometric picture of an underlying schematic structure.  The
variation with mass ratio is substantial. The two remaining states for
moderate and large mass ratios appears to be consistent with the light
particles at roughly twice the distance between light and heavy
particles.  For small mass ratios we find for both ground and excited
states that the light particle is located at distances much larger
than the distance between the two heavy particles

The disappearance of excited states into the continuum for threshold
values of the mass ratio is related to specific bound-state structures
close to the thresholds.  The mean-square radii here prove themselves
to be efficient measures of the corresponding structures.  We show
numerically that these mean-square radii diverge inversely
proportional to the deviation of the three-body energy from the
two-body threshold energy.  Comparing distances between identical and
unequal particles we conclude that the threshold structure corresponds
to one of the identical particles far away from a bound dimer of the
other two particles.  Thus, this far-away particle is being ejected
into the continuum.  It should be possible to confirm these predictions
by cold atom experiments where three-body atomic systems of
$^6$Li-$^{133}$Cs-$^{133}$Cs and $^6$Li-$^{87}$Rb-$^{87}$Rb among
others, are already produced in laboratories.

The technique we developed is based on the schematic, but hugely
popular, zero-range interaction treated in momentum space.  This
interaction is the extreme limit of a short-range potential, and
therefore in general the simplest tool to provide quantitative
information about universal properties.  The present method to
calculate mean-square radii is a special application to derive
structure information through expectation values.  The method is
simpler and faster than using brute force to obtain wave functions
which afterwards supply the desired observables. An interesting 
continuation of the present study is to extend it to more than 
three atoms and see how the universal relations derived in 
\cite{hammerprl2004} would change for a mass-imbalanced system.

The method can be used directly to investigate the dependence of radii
on the strengths of the two-body interactions.  The formalism can be
relatively easily generalized to apply to three-body systems where one
or more two-body subsystems are unbound, whereas the three-body system
still remains bound.  All applications may be extended to three
distinguishable particles with their different masses and interaction
parameters.  The application to two spatial dimensions is interesting
because this is a rigorous limit with unique properties compared to
other geometries.  However, also both one and three dimensions are equally 
accessible, and a continuous variation between dimensions also becomes
increasingly interesting.

\acknowledgments
This work was partly supported by funds provided by the Brazilian agencies 
FAPESP (2016/01816-2), CNPq and CAPES (88881.030363/2013-01).




\begin{thebibliography}{28}

\bibitem{pethick} C. J. Pethick and H. Smith, {\it Bose-Einstein Condensation in 
Dilute Gases} (Cambridge University Press 2008), Chap. 5.

\bibitem{modugnoPRA03} G. Modugno, F. Ferlaino, R. Heidemann, G. Roati and M. Inguscio, 
Phys. Rev. A {\bf 68}, 011601 (2003).

\bibitem{gunterPRL05} K. G\"unter, T. St\"oferle, H. Moritz, M. K\"ohl and T. Esslinger, 
Phys. Rev. Lett. {\bf 95}, 230401 (2005).

\bibitem{castinpra2012} F. Werner and Y. Castin, Phys. Rev. A {\bf 86}, 053633 (2012).

\bibitem{bellottifbs2015} F. F. Bellotti and M. T. Yamashita, Few-Body Syst. 
{\bf 56}, 905 (2015).

\bibitem{jensenrmp2004} A. S. Jensen, K. Riisager, D. V. Fedorov and E. Garrido, 
Rev. Mod. Phys. {\bf 76}, 215 (2004).

\bibitem{braatenpr2006} E. Braaten and H. W. Hammer, Phys. Rep. {\bf 428}, 259 (2006).

\bibitem{fre12} T. Frederico, L. Tomio, A. Delfino, M. R. Hadizadeh and 
M. T. Yamashita, Few-Body Syst. {\bf 51}, 87 (2011).

\bibitem{kunitskiscience2015} M. Kunitski et al., Science {\bf 348}, 551 (2015).

\bibitem{cencekjcp2012} W. Cencek et al., J. Chem. Phys. {\bf 136}, 224303 (2012).

\bibitem{kolganovafbs2011} E. A. Kolganova, A. K. Motovilov and W. Sandhas, 
Few-Body Syst. {\bf 51}, 249 (2011).
 
\bibitem{skorniakov1957} G. V. Skornyakov and K. A. Ter-Martirosyan, 
Sov. Phys. JETP {\bf 4}, 648 (1957).

\bibitem{efimov}  V. Efimov, Phys. Lett. B {\bf 33}, 563 (1970); V. Efimov, ,
Nucl. Phys. A {\bf 362}, 45 (1981).

\bibitem{nielsenpr2001} E. Nielsen, D. V. Fedorov, A. S. Jensen and E. Garrido, 
Phys. Rep. {\bf 347}, 373 (2001).


\bibitem{yama15} M. T. Yamashita, F. F. Bellotti, T. Frederico,  D. V. Fedorov, 
A. S. Jensen and N. T. Zinner, J. Phys. B {\bf 48}, 025302 (2015).

\bibitem{bruchpra1979} L. W. Bruch and J. A. Tjon, Phys. Rev. A {\bf 19}, 425 (1979).

\bibitem{adhikarijp1986} S. K. Adhikari, Am. J. Phys. {\bf 54}, 362 (1986).

\bibitem{adhikaripra1993} S. K. Adhikari, A. Delfino, T. Frederico and L. Tomio, 
Phys. Rev. A {\bf 47}, 1093 (1993).

\bibitem{nie99} E. Nielsen, D.V. Fedorov and A.S. Jensen, 
Few-Body Systems {\bf 27}, 15 (1999).
 
\bibitem{bellottijpb2011} F. F. Bellotti, T. Frederico, M. T. Yamashita, D. V. Fedorov, 
A. S. Jensen and N. T. Zinner, J. Phys. B {\bf 44}, 205302 (2011).

\bibitem{bellottipra2012} F. F. Bellotti, T. Frederico, M. T. Yamashita, D. V. Fedorov, 
A. S. Jensen and N. T. Zinner, Phys. Rev. A {\bf 85}, 025601 (2012).

\bibitem{bellottijpb2013} F. F. Bellotti, T. Frederico, M. T. Yamashita, D. V. Fedorov,
A. S. Jensen and N. T. Zinner, J. Phys. B {\bf 46}, 055301 (2013).

\bibitem{khurifbs2002} N. N. Khuri, A. Martin and T.-T. Wu, Few-Body Syst. {\bf 31}, 
83 (2002).

\bibitem{levinsenprx2014} J. Levinsen, P. Massignan and M. M. Parish, Phys. Rev. X
{\bf 4}, 031020 (2014).

\bibitem{esry2014} J. P. D'Incao and B. D. Esry, Phys. Rev. A {\bf 90}, 042707 (2014).

\bibitem{esry2015} J. P. D'Incao, F. Anis and B. D. Esry, Phys. Rev. A {\bf 91}, 
062710 (2015).

\bibitem{limZPA80} T. K. Lim and B. Shimer, Z. Phys. A {\bf 297}, 185 (1980).

\bibitem{yamashitapra2003} M. T. Yamashita, R. S. Marques de Carvalho, L. Tomio 
and T. Frederico, Phys. Rev. A {\bf 68}, 012506 (2003).

\bibitem{hammerprl2004} H.-W. Hammer and D. T. Son, Phys. Rev. Lett. {\bf 93}, 
250408 (2004).

\bibitem{fonsecaNPA78} A. C. Fonseca, E. F. Redish and P. E. Shanley, Nucl. 
Phys. A {\bf 320}, 273 (1979).

\bibitem{brunnian} M. T. Yamashita, D. V. Fedorov and A. S. Jensen, Phys. Rev. A {\bf 81}, 
063607 (2010).

\end{thebibliography}
\end{document}